\documentclass[aps,prl,preprint,groupedaddress,showpacs]{revtex4}
\usepackage{graphicx}
\usepackage{dcolumn}
\usepackage{bm}
\input epsf
\epsfclipon

\begin{document}
\title{Mean-field theory for clustering coefficients in Barab\'{a}si-Albert networks}
\author{Agata Fronczak, Piotr Fronczak and Janusz A. Ho\l yst}
\affiliation{Faculty of Physics and Center of Excellence for
Complex Systems Research, Warsaw University of Technology,
Koszykowa 75, PL-00-662 Warsaw, Poland}
\date{\today}
\begin{abstract}
We applied a mean field approach to study clustering coefficients
in Barab\'{a}si-Albert networks. We found that the local
clustering in BA networks depends on the node degree. Analytic
results have been compared to extensive numerical simulations
finding a very good agreement for nodes with low degrees.
Clustering coefficient of a whole network calculated from our
approach perfectly fits numerical data.
\end{abstract} \pacs{89.75.-k, 02.50.-r, 05.50.+q} \maketitle

{\it Introduction.} During the last decade networks became a very
popular research domain among physicists (for a review see
\cite{0_handbook,0_dorogov,BAprzeglad}). It is not surprising,
since networks are everywhere. They surround us. In our daily life
we participate in dozens of them. A number of social institutions,
communication and biological systems may be represented as
networks i.e. sets of nodes connected by links. It was observed
that despite functional diversity most of real web-like systems
share similar structural properties. The properties are:
fat-tailed degree distribution (that allows for hubs i.e. nodes
with high degree), small average distance between any two nodes
(the so-called {\it small world} effect) and a large penchant for
creating cliques (i.e. highly interconnected groups of nodes).

A number of network construction procedures have been proposed to
incorporate the characteristics. The Barab\'{a}si-Albert (BA)
\cite{BA_science,BA_physicaa} growing network model is probably
the best known. Two important ingredients of the model are:
continuous network growth and preferential attachment. The network
starts to grow from an initial cluster of $m$ fully connected
sites. Each new node that is added to the network creates $m$
links that connect it to previously added nodes. The preferential
attachment means that the probability of a new link to end up in a
vertex $i$ is proportional to the connectivity $k_{i}$ of this
vertex
\begin{equation}\label{PAR}
\Pi_{i}=m\frac{k_{i}}{\sum_{j}k_{j}}.
\end{equation}
Taking into account that $\sum_{j}k_{j}=2mt$ the last formula may
be rewritten as $\Pi_{i}=k_{i}/(2t)$. By means of mean field
approximation \cite{BA_physicaa} one can find that the average
degree of a node $i$ that entered the network at time $t_{i}$
increases with time as a power-law
\begin{equation}\label{ki}
k_{i}(t)=m\sqrt{\frac{t}{t_{i}}}.
\end{equation}
Taking advantage of the above formula one can calculate the
probability that two randomly selected nodes $i$ and $j$ are
nearest neighbors. It is given by
\begin{equation}\label{pij}
p_{ij}=\frac{m}{2}\frac{1}{\sqrt{t_{i}t_{j}}}.
\end{equation}
It was shown that the degree distribution in BA network follows a
power-law
\begin{equation}\label{pk}
P(k)=\frac{2m^{2}}{k^{3}},
\end{equation}
where $k=m,m+1,\dots,m\sqrt{t}$. The power law degree distribution
is characteristic of many real-world networks and the scaling
exponent $\alpha_{BA}=3$ is close to those observed in real
systems $\alpha_{real}\in(2,3)$. It was also shown that the BA
model is a small world. The mean distance between sites in the
network having $t$ nodes behaves as $l\sim\ln t/\ln\ln t$
\cite{fronczak,havlin}. The only shortcoming of the model is that
it does not incorporate a high degree of cliqueness observed in
real networks.

In this paper we study cliqueness effects in BA networks. The
cliqueness is measured by the clustering coefficient $C$
\cite{watts,fronczak1}. The clustering coefficient $C_{i}$ of a
single node $i$ describes the density of connections in the
neighborhood of this node. It is given by the ratio of the number
$E_{i}$ of links between the nearest neighbors of $i$ and the
potential number of such links $E_{max}=k_{i}(k_{i}-1)/2$
\begin{equation}\label{ci1}
C_{i}=\frac{E_{i}}{E_{max}}=\frac{2E_{i}}{k_{i}(k_{i}-1)}.
\end{equation}
The clustering coefficient $C$ of the whole network is the average
of all individual $C_{i}$'s. It is known, from numerical
calculations, that the clustering coefficient in BA networks
rapidly decreases with the network size $t$. In this article we
apply a mean field approach to study the parameter. Our
calculations confirm that in the limit of large ($t\gg 1$) and
dense ($m\gg 1$) BA networks the clustering coefficient scales as
the clustering coefficient in random graphs
\cite{klemm,kertesz,newman} with an appropriate scale-free degree
distribution (\ref{pk})
\begin{equation}\label{Crg}
C=\frac{(m-1)}{8}\frac{(\ln t)^{2}}{t}.
\end{equation}
We also show that the individual clustering coefficient $C_{i}$ in
BA network weakly depends on node's degree $k_{i}$. The dependence
is almost invisible when one looks at numerical data presented by
other authors \cite{ravasz}.

\begin{figure}\epsfxsize=10cm\epsfbox{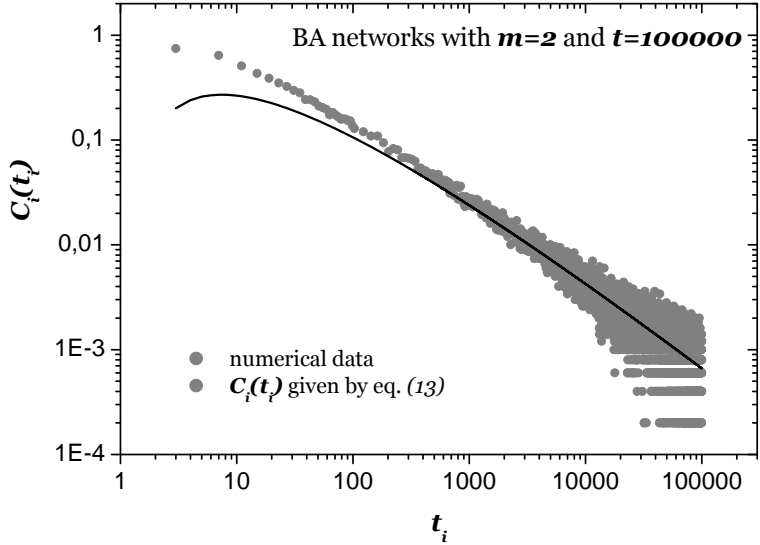}
\caption{The initial value of the local clustering coefficient
$C_{i}(t_{i})$ (averaged over $1000$ BA networks).}
\label{figciti}
\end{figure}

{\it Mean field approach.} Let us concentrate on a certain node
$i$ in a BA network of size $t$. We assume that $m\geq 2$. The
case of $m=1$ is trivial. BA networks with $m=1$ are trees thus
the clustering coefficient in these networks is equal to zero. By
the definition (\ref{ci1}) the clustering coefficient $C_{i}$
depends on two variables $E_{i}$ and $k_{i}$. Since in the BA
model only new nodes may create links the coefficient $C_{i}$
changes only when its degree $k_{i}$ changes i.e. when new nodes
create connections to $i$ and $x\in\langle 0,m-1\rangle$ of its
nearest neighbors. The appropriate equation for changes of $C_{i}$
is then
\begin{equation}\label{eq1}
\frac{dC_{i}}{dt}=\sum_{x=0}^{m-1}\widetilde{p}_{ix}\Delta
C_{ix}\:,
\end{equation}
where $\Delta C_{ix}$ denotes the change of the clustering
coefficient when a new node connects to the node $i$ and to $x$ of
the first neighbors of $i$, whereas $\widetilde{p}_{ix}$ describes
the probability of this event. $\Delta C_{ix}$ is simply the
difference between clustering coefficients of the same node $i$
calculated after and before a new node attachment
\begin{equation}\label{eq1a}
\Delta
C_{ix}=\frac{2(E_{i}+x)}{k_{i}(k_{i}+1)}-\frac{2E_{i}}{k_{i}(k_{i}-1)}=
-\frac{2C_{i}}{k_{i}+1}+\frac{2x}{k_{i}(k_{i}+1)}.
\end{equation}
The probability $\widetilde{p}_{ix}$ is a product of two factors.
The first factor is the probability of a new link to end up in
$i$. The probability is given by (\ref{PAR}). The second one is
the probability that among the rest of $(m-1)$ new links $x$ links
connect to nearest neighbors of $i$. It is equivalent to the
probability that $(m-1)$ Bernoulli trials with the probability for
success equal to
$\sum_{j*}k_{j}/\sum_{v}k_{v}=\sum_{j*}k_{j}/(2mt)$ result in $x$
successes ($\sum_{j*}$ runs over the nearest neighbors of the node
$i$). Replacing the sum $\sum_{j*}$ by an integral one obtains
\begin{equation}
\sum_{j*}k_{j}= \int_{1}^{t}k_{j}\:p_{ij}\:dt_{j}=
\frac{m}{2}\:k_{i}\ln t.
\end{equation}
Summarizing the above discussion one yields the relation
\begin{equation}\label{eq1b}
\widetilde{p}_{ix} = \frac{k_{i}}{2t}
\left(^{m-1}_{\:\:\:\:x}\right) \left(\frac{k_{i}\ln
t}{4t}\right)^{x}\left(1-\frac{k_{i}\ln t}{4t}\right)^{m-1-x}
\end{equation}
Now, inserting (\ref{ki}), (\ref{eq1a}) and (\ref{eq1b}) into
(\ref{eq1}) one obtains after some algebra
\begin{equation}\label{eq2}
\frac{dC_{i}}{dt}=-\frac{m}{(mt+\sqrt{tt_{i}})}C_{i}+
\frac{m(m-1)\ln t}{4(mt^{2}+t\sqrt{tt_{i}})}.
\end{equation}
Solving the equation for $C_{i}$ one gets
\begin{equation}\label{cit}
C_{i}(t)=\frac{(m-1)}{8(\sqrt{t}+\sqrt{t_{i}}/m)^{2}} \left((\ln t
)^{2}-\frac{4}{m}\sqrt{\frac{t_{t}}{t}}\ln
t-\frac{8}{m}\sqrt{\frac{t_{i}}{t}}+B\right),
\end{equation}
where $B$ is an integration constant and may be determined from
the initial condition $C_{i}(t_{i})$ that describes the clustering
coefficient of the node $i$ exactly at the moment of its
attachment $t_{i}$
\begin{equation}\label{citi}
C_{i}(t_{i})=\frac{1}{2}\sum_{j}\sum_{v}p_{ij}\:p_{iv}\:p_{jv}\:/\left(^{m}_{\:2}\right)
=\frac{m^{2}}{8(m-1)}\frac{(\ln t_{i})^{2}}{t_{i}}.
\end{equation}
Following the notation introduced by Bianconi and Capocci
\cite{bianconi}, the initial clustering coefficient $C_{i}(t_{i})$
may may be written as
\begin{equation}
C_{i}(t_{i})=\frac{1}{\left(^{m}_{\:2}\right)}\left[\frac{\partial\langle
N_{3}(t) \rangle}{\partial t}\right]_{t=t_{i}},
\end{equation}
where $\partial\langle N_{3}(t) \rangle/\partial t$ describes how
the number of triangular loops increases in time.
Fig.\ref{figciti} shows the prediction of the equation
(\ref{citi}) in comparison with numerical results. For small
values of $t_{i}$ the numerical data differ from the theory in a
significant way. This can be explained by the fact that the
formula for the probability of a connection $p_{ij}$ (\ref{pij}),
that we use three times in (\ref{citi}), holds only in the
asymptotic region $t_{i}\longrightarrow\infty$.

Taking into account the initial condition $C_{i}(t_{i})$ and
neglecting mutually compensating terms that occure in (\ref{cit})
after putting $B$ calculated from (\ref{citi}) one obtains the
formula for time evolution of the clustering coefficient of a
given node $i$
\begin{equation}\label{citt}
C_{i}(t)=\frac{(m-1)}{8(\sqrt{t}+\sqrt{t_{i}}/m)^{2}} \left((\ln t
)^{2}+\frac{4m}{(m-1)^{2}}(\ln t_{i})^{2}\right).
\end{equation}
\begin{figure}\epsfxsize=10cm\epsfbox{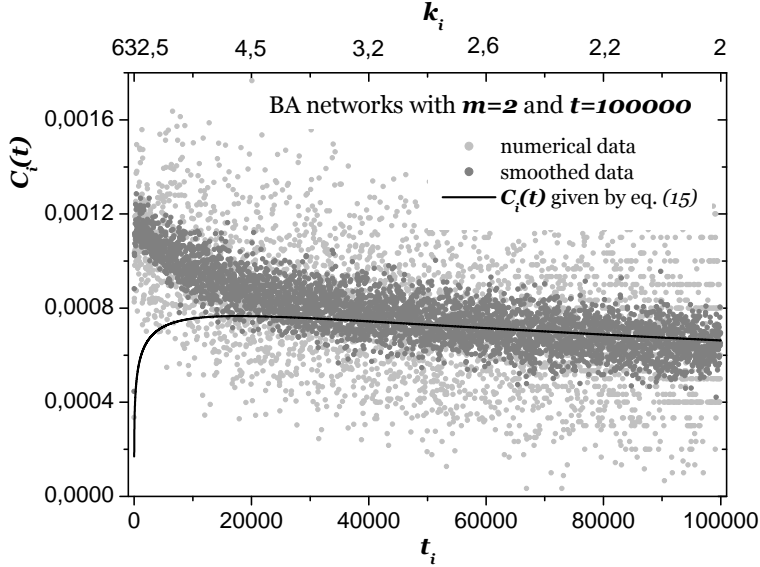}
\caption{The local clustering coefficient $C_{i}(t)$ as a function
$t_{i}$ (averaged over $10^{4}$ networks). Note that the $k_{i}$
axis is nonlinear.} \label{figcit}
\end{figure}
Let us note that if $t_{i}\ll t$ or $m\gg 1$ the local clustering
coefficient does not depend on the node under consideration and
approaches $C_{i}(t)\simeq (m-1)(\ln t)^{2}/(8t)$ i.e. the formula
(\ref{Crg}) that gives the the clustering coefficient of a random
graph with a power-low degree distribution (\ref{pk}). Since one
knows how the node's degree evolves in time (\ref{ki}) one can
also calculate the formula for $C_{i}(k_{i})$. At the
Fig.\ref{figcit} we present the formula (\ref{citt}) (solid line)
and corresponding numerical data (scatter plots). The two kinds of
scatter plots represent respectively: real data (light gray
circles) and the same data subjected adjacent averaging smoothing
(dark gray circles). As before (see Fig.(\ref{figciti})), we
observe a significant disagreement between the numerical data and
the theory for small $t_{i}$. We suspect that the effect has the
same origin i.e. the relations (\ref{ki}) and (\ref{pij}) that we
use in our derivation work well only in the asymptotic region
$t_{i}<t\longrightarrow\infty$.

\begin{figure}\epsfxsize=10cm\epsfbox{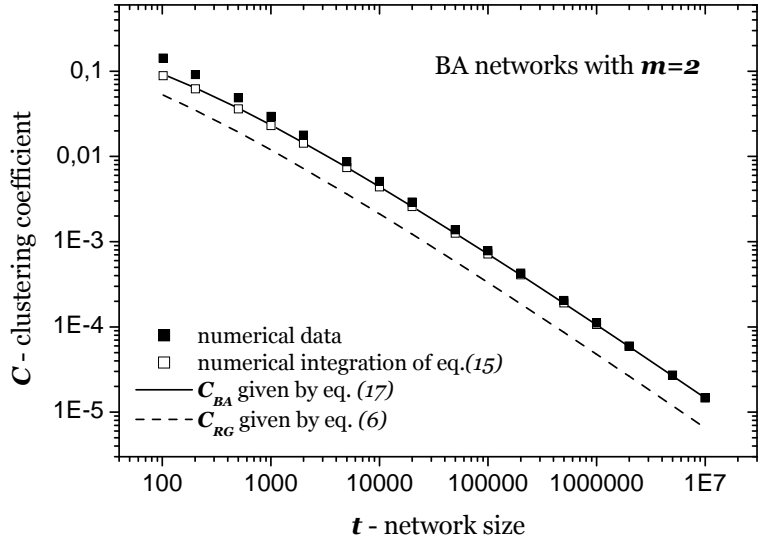}
\caption{The clustering coefficient $C$ of a whole BA network as a
function of the network size $t$ (averaged over $100$ networks).}
\label{figcn}
\end{figure}

To obtain the clustering coefficient $C$ of the whole network the
expression (\ref{citt}) has to be averaged over all nodes within a
network $C=\int_{1}^{t}C_{i}(t)dt_{i}/t$. We were not able to find
an exact analytic form of this integral but corresponding
numerical values (open squares at the Fig.\ref{figcn}) fit very
well a mean field approximation that we propose below (solid line
at the Fig.\ref{figcn})
\begin{equation}\label{cmfa}
C=\left\langle C_{i}(t)\right\rangle_{i}\simeq \frac{(m-1)}{8}
\frac{\left\langle\:(\ln t)^{2}+\frac{4m}{(m-1)^{2}}(\ln
t_{i})^{2}\:\right\rangle_{i}}{\left\langle\:(\sqrt{t}+\sqrt{t_{i}}/m)^{2}\:\right\rangle_{i}}.
\end{equation}
After performing separate integration of the numerator and the
denominator one gets
\begin{equation}\label{c}
C=\frac{6m^{2}\left((m+1)^{2}(\ln t)^2-8m\ln
t+8m\right)}{8(m-1)(6m^{2}+8m+3)t}.
\end{equation}
For large ($t\longrightarrow\infty$) and dense ($m\gg 1$) networks
the above formula approaches (\ref{Crg}). The effect lets us
deduce that the structural correlations \cite{dorogov}
characteristic for growing BA networks become less important in
larger and denser networks. The same was suggested in
\cite{fronczak}. Fig.\ref{figcn} shows the average clustering
coefficient in BA networks as a function of the network size $t$
compared with the analytical formula (\ref{c}).

{\it Conclusions.} In summary, we applied a mean field approach to
study clustering effects in Barab\'{a}si-Albert networks. We found
that the BA networks do not show the homogeneous clustering as
suggested in \cite{klemm,ravasz}. We derived a general formula for
the clustering coefficient $C$ characterizing the whole BA
network. We found that in the limit of large
($t\longrightarrow\infty$) and dense ($m\gg 1$) networks both the
local ($C_{i}$) and the global ($C$) clustering coefficients
approach clustering coefficient derived for a random graph with a
power-low degree distribution (\ref{pk}). Our derivations were
checked against numerical simulation of BA networks finding a very
good agreement.

{\it Acknowledgments.} We would like to thank Jaros\l aw Suszek
and Daniel Kiko\l a for their help in computer simulations. AF
thanks The State Committee for Scientific Research in Poland for
support under grant No. $2 P03B 013 23$. The work of JAH was
supported by the special program of the Warsaw University of
Technology {\it Dynamics of Complex Systems}.


\end{document}